\documentclass[twocolumn,aps]{revtex4}

\newcommand {\bc}{\begin{center}}
\newcommand {\ec}{\end{center}}
\newcommand {\bea}{\begin{eqnarray}}
\newcommand {\eea}{\end{eqnarray}}
\newcommand {\be}{\begin{equation}}
\newcommand {\ee}{\end{equation}}

\def\lsim{\mathrel{\rlap{\lower4pt\hbox{\hskip1pt$\sim$}}
    \raise1pt\hbox{$<$}}}
\def\gsim{\mathrel{\rlap{\lower4pt\hbox{\hskip1pt$\sim$}}
    \raise1pt\hbox{$>$}}}
\usepackage{graphicx}
\usepackage{color}
\usepackage{soul}

\begin{document}


\title{Model-independent determination of the shear viscosity of 
a trapped unitary Fermi gas: Application to high temperature data}

\author{M. Bluhm and T.~Sch\"afer}

\affiliation{Department of Physics, North Carolina State University,
Raleigh, NC 27695}

\begin{abstract}
Determinations of the shear viscosity of trapped ultracold gases
suffer from systematic, uncontrolled uncertainties related to the 
treatment of the dilute part of the gas cloud. In this work we present 
an analysis of expansion experiments based on a new method, 
anisotropic fluid dynamics, that interpolates between Navier-Stokes
fluid dynamics at the center of the cloud and ballistic behavior 
in the dilute corona. We validate the method using a comparison
between anisotropic fluid dynamics and numerical solutions of the 
Boltzmann equation. We then apply anisotropic fluid dynamics 
to the expansion data reported by Cao et al. In the high temperature
limit we find $\eta=0.282(mT)^{3/2}$, which agrees within about 5\% with 
the theoretical prediction  $\eta=0.269(mT)^{3/2}$.
\end{abstract}

\maketitle

{\it Introduction:}
 A number of studies have been devoted to extracting the transport
properties of dilute atomic Fermi gases. Quantities of interest
include the shear viscosity \cite{Kinast:2005,Schafer:2007pr,Turlapov:2007,Cao:2010wa,Vogt:2011np,Schafer:2011my,Enss:2012nx,Elliott:2013b,Bluhm:2014uza,Joseph:2014,Brewer:2015hua},
the bulk viscosity \cite{Elliott:2013}, and the spin diffusion constant
\cite{Sommer:2011,Bruun:2011b,Koschorreck:2013}. These transport coefficients
provide valuable information about the nature of the low energy degrees 
of freedom. Strongly correlated Fermi gases also contribute important 
insights into the transport properties of other quantum many-body systems, 
such as high-$T_c$ superconductors or the quark-gluon plasma 
\cite{Guo:2010,Schafer:2009dj,Adams:2012th}. Truly model-independent 
determinations of the transport coefficients of trapped atomic gases 
have so far been precluded, however, by the fact that there is a 
transition from fluid dynamical behavior in the dense part of the 
cloud to weakly collisional kinetic behavior in the dilute corona. 

 Consider, for example, a unitary Fermi gas expanding after release 
from a deformed harmonic trap~\cite{Schaefer:2009px}. Fluid dynamics
predicts that the difference in pressure gradients along the short 
and the long axis of the cloud translates into a larger acceleration 
along the short direction. This implies that the aspect ratio $A_R$ 
of the cloud, which is initially much smaller than one, quickly grows 
and eventually exceeds unity, as was first observed by O'Hara et 
al.~\cite{OHara:2002}. Shear viscosity $\eta$ slows down the acceleration 
in the transverse direction, and measurements of $A_R(t)$ for 
different initial values of $T/T_F$, where $T_F$ is the Fermi 
temperature, can be used to constrain the dependence of $\eta(n,T)$ on 
density $n$ and temperature $T$. This task is simplified by the scale 
invariance of the unitary Fermi gas, which implies that the bulk 
viscosity vanishes, and that $\eta=(mT)^{3/2}f(n/(mT)^{3/2})$, where 
$f(x)$ is a universal function. Note that we use units $\hbar=k_B=1$. 

 The natural tool for extracting $\eta(n,T)$ is the Navier-Stokes
(NS) equation. The problem in determining $\eta(n,T)$ is that $A_R(t)$ 
is a global property of the cloud, and that the NS equation breaks 
down in the dilute corona, where the mean free path is large compared 
to the density and the pressure scale heights. Because the total number 
of particles in the corona is small, one might hope that this does not lead 
to serious difficulties. Unfortunately, this is not the case: The rate 
of dissipative heating is $\dot{q}=\frac{\eta}{2}(\sigma_{ij})^2$, where 
$\sigma_{ij}= \nabla_iu_j+ \nabla_j u_i-\frac{2}{3}\delta_{ij} \vec{\nabla}
\cdot\vec{u}$ is the strain tensor, and $\vec{u}$ is the fluid velocity. 
In the dilute limit kinetic theory predicts that the shear viscosity is 
only a function of temperature, and not of density, $\eta\sim (mT)^{3/2}$ 
\cite{Bruun:2005,Bruun:2006}. The square of the strain tensor scales as 
$(\sigma_{ij})^2\sim \tau_{\it exp}^{-2}$, where $\tau_{\it exp}^{-1}=\vec{\nabla}
\cdot \vec{u}$ is the expansion rate of the fluid. This means that the local 
heating rate is $\dot{q}\sim T^{3/2}\tau_{\it exp}^{-2} \sim T^3$, independent 
of density \footnote{
This estimate is based on properties of the scaling solution of the Euler 
equation for an expanding gas. The velocity field is linear $u_i=\alpha_i(t)
x_i$, where $\alpha_i=\dot{b}_i(t)/b_i(t)$ and $b_i$ is the scale factor of 
the expansion in the $i$ direction. The density of a co-moving fluid element 
scales as $n\sim 1/b_\perp^2$, and the temperature scales as $T\sim n^{2/3}$. 
Finally, after the initial acceleration period we have $b_i\sim \omega_i t$ 
and $\alpha_i \sim \omega_i/b_i$.}.
Thus, integrating the NS equation over volume leads to the prediction 
that dissipation produces an infinite amount of heat. This result is, of 
course, an artifact of applying the NS equation in a regime where the mean 
free path is large. It implies, however, that any attempt to address this 
problem by imposing a cutoff radius will give results that are very sensitive 
to the precise nature of the cutoff.

{\it Prior work:}
 Previous analyses have dealt with this issue in a variety of ways. In 
\cite{Schafer:2007pr} it was argued that collective mode and expansion
experiments primarily constrain the trap integral of the shear viscosity, 
$\alpha_n\equiv\frac{1}{N}\int d^3x\, \eta(n_0(\vec{x}),T_0)$, where $N$ 
is the total number of particles, $n_0(\vec{x})$ is the initial density, and 
$T_0$ is the initial temperature. The integration volume was restricted to 
lie within the surface of last scattering, defined using the mean free path 
computed in kinetic theory. Later, Cao et al. \cite{Cao:2010wa} assumed that 
the local shear viscosity scales as $\eta(\vec{x})=n(\vec{x})[\eta(0)/n(0)]$, 
so that $\alpha_n=\eta(0)/n(0)$ is determined by $\eta$ and $n$ at the trap 
center. This assumption has a number of nice properties, because for a 
scaling expansion $\eta(0)/n(0)$ is approximately independent of time. 
In the more recent work by Joseph et al.~\cite{Joseph:2014} the integration 
volume was restricted to the interior of an ellipsoid. The length of the 
principle axes was taken to be $R_i=\gamma\langle x_i^2\rangle^{1/2}$, where 
$\langle x_i^2\rangle^{1/2}$ is the the rms radius, and $\gamma$ is a 
temperature-independent coefficient that was fitted in order to reproduce 
the theoretically computed high-$T$ limit of the shear viscosity, $\eta=
\frac{15}{32\sqrt{\pi}}(mT)^{3/2}$  \cite{Bruun:2005,Bruun:2006}.

\begin{figure}[t]
\bc\includegraphics[width=7.5cm]{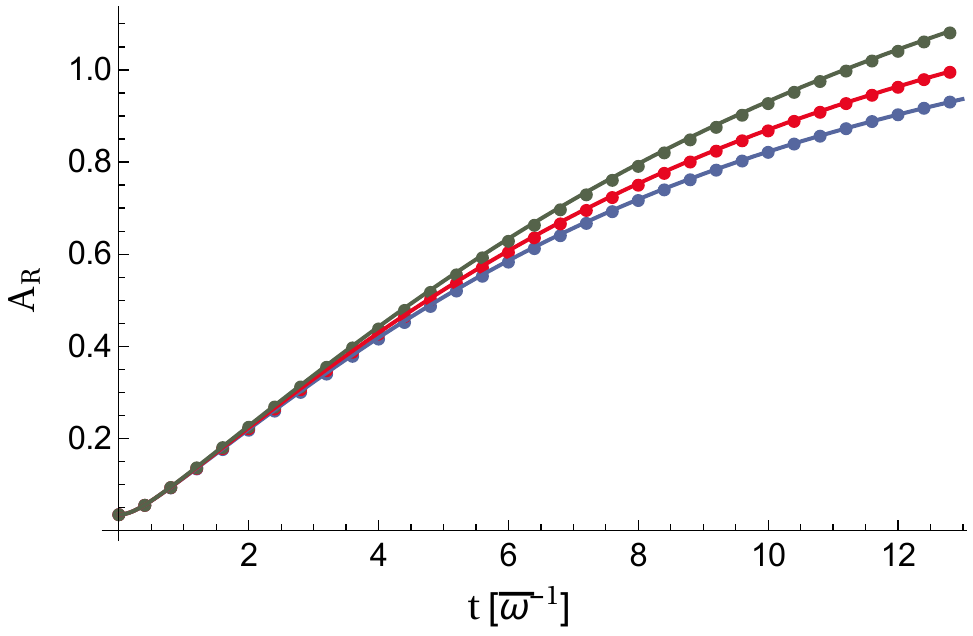}\ec
\caption{\label{fig_comp}
 This figure shows the aspect ratio $A_R$ of an expanding unitary 
Fermi gas as a function of time $t$ in units of the inverse mean trap 
frequency $\bar{\omega}^{-1}$. The three curves correspond to three 
different initial temperatures $T/T_F=0.79,1.11,1.54$ (from top to bottom).
The solid lines show results obtained using anisotropic fluid dynamics, 
and the points are solutions of the Boltzmann equation obtained by Pantel 
et al.~\cite{Pantel:2014jfa}. The aspect ratio is defined using the rms 
radii, $A_R=[\langle x^2\rangle/\langle z^2\rangle]^{1/2}$. }
\end{figure}

{\it Anisotropic fluid dynamics:}
 These methods are clearly not fully satisfactory, because they involve
model assumptions for which the error cannot be quantified. For example, 
the analysis of Cao et al.~gives $\eta=0.33(mT)^{3/2}$ in the high temperature 
limit $T\gg T_F$~\cite{Cao:2010wa}. This agrees to within 25\% with the 
theoretical prediction, but there is no a priori estimate of the theoretical 
error related to the assumption $\eta(\vec{x})=n(\vec{x})[\eta(0)/n(0)]$. 
Also, there is no reliable method for estimating the systematic uncertainty 
in the low temperature data obtained by Joseph et al.~\cite{Joseph:2014}.

A possible approach that resolves the difficulty is to couple a fluid 
dynamical calculation for the center of the cloud with a kinetic treatment 
based on the Boltzmann equation for the dilute corona. However, this method 
is computationally very demanding, and extensive studies would be required 
to establish that the results are independent of the prescription for 
switching between thermodynamic variables in fluid dynamics and distribution 
functions in kinetic theory. A much simpler approach, termed anisotropic 
fluid dynamics, was recently proposed in \cite{Bluhm:2015raa}. This method 
has also been studied in connection with relativistic heavy-ion collisions
\cite{Florkowski:2010,Martinez:2010sc}. The idea is to include certain
non-hydrodynamic variables in the fluid dynamical description. In the 
limit of short mean free paths, these variables relax to their equilibrium
values on a microscopic time scale, and NS theory is recovered. In the 
limit of long mean free paths, in contrast, the non-hydrodynamic modes are 
approximately conserved, and the additional conservation laws ensure a smooth
transition to free streaming. 

 The fluid dynamical variables describing a non-relativistic fluid in 
the normal phase are the mass density $\rho$, the momentum density 
$\vec{\pi}=\rho\vec{u}$, and the energy density ${\cal E}$. The 
conservation laws can be written as 
\bea
\label{rho_lag}
 D_0\rho &=& -\rho \vec{\nabla}\cdot \vec{u}\, , 
                        \frac{\mbox{}}{\mbox{}} \\
\label{u_lag}
 D_0 u_i & = & - \frac{1}{\rho} \left( \nabla_i P 
 + \nabla_j \delta \Pi_{ij} \right) \, , \\
\label{e_lag}
 D_0 \epsilon & = & - \frac{1}{\rho} \nabla_i \left( u_i P 
 + \delta \jmath^{\cal E}_i \right) \, , 
\eea
where we have introduced the comoving time derivative $D_0=\partial_0 +\vec{u}
\cdot\vec{\nabla}$, the energy per mass $\epsilon={\cal E}/\rho$, and the 
pressure $P$. In NS theory the dissipative stress tensor is given by $\delta
\Pi_{ij}=-\eta\sigma_{ij}$ and the dissipative energy current is $\delta 
\jmath^{\cal E}_i = u_j\delta \Pi_{ij}$. For simplicity, we neglect the effects 
of heat conduction, which are not important for the physical systems studied 
in this work \cite{Schafer:2010dv}. The fluid dynamical equations close once 
we provide an equation of state $P=P({\cal E}^0)$, where ${\cal E}^0={\cal E}
-\frac{1}{2}\rho\vec{u}^2$ is the energy density in the fluid rest frame. 
The unitary Fermi gas is scale invariant and $P=\frac{2}{3}{\cal E}^0$. 

 In anisotropic fluid dynamics we treat the components of the dissipative 
stress tensor as additional, independent, fluid dynamical variables. In the 
present case the stresses remain diagonal and we only have to keep the 
diagonal components of $\delta\Pi_{ij}$ \footnote{The stresses are diagonal
because of the symmetries of the trapping potential. The Euler equation
implies $\partial_0 u_i \sim \nabla_i P \sim x_i$ so that $\nabla_i u_j =0$ 
for $i\neq j$. This feature is preserved by viscous corrections.}. 
We define anisotropic components of the pressure, $P_a$ for $a=1,2,3$, 
and write $\delta\Pi_{ij}=\sum_a\delta_{ia}\delta_{ja}\Delta P_a$, where 
$\Delta P_a=P_a-P$. We also define anisotropic components of the energy 
density ${\cal E}_a$ such that ${\cal E}=\sum_a{\cal E}_a$. The anisotropic 
components of the energy per mass satisfy the fluid dynamical equations 
\cite{Bluhm:2015raa}
\be 
\label{e_a_lag}
 D_0 \epsilon_a = - \frac{1}{\rho}
 \nabla_i \left[ \delta_{ia} u_i P + (\delta \jmath^{\cal E}_a)_i \right] 
 - \frac{P}{2\eta\rho} \,\Delta P_a\, , 
\ee
where $\epsilon_a={\cal E}_a/\rho$ and $(\delta \jmath^{\cal E}_a)_i 
= \delta_{ia} u_j \delta \Pi_{ij}$. To close the fluid dynamical equations 
we have to provide an equation of state. For a scale invariant fluid we 
have $P_a({\cal E}^0_a)=2\,{\cal E}^0_a$ with ${\cal E}^0_a={\cal E}_a-
\frac{1}{2}\rho u_a^2$. Then $P=\frac{1}{3}\sum_a P_a$ satisfies the 
isotropic equation of state and equ.~(\ref{e_a_lag}) reproduces the 
equation of energy conservation equ.~(\ref{e_lag}) when summed 
over $a$. Equations (\ref{rho_lag})-(\ref{e_a_lag}) can be solved using 
standard techniques in computational fluid dynamics. We have developed
a code based on the PPM scheme of Colella and Woodward 
\cite{Colella:1984,Blondin:1993,Schafer:2010dv,Bluhm:2015raa}.

 The precise form of the fluid dynamical equations (\ref{rho_lag}-
\ref{e_a_lag}) can be derived using moments of the Boltzmann equation
\cite{Bluhm:2015raa}. In particular, the new equation~(\ref{e_a_lag}) 
arises from taking moments with $p_a^2/(2m)$, where $p_a$ is a Cartesian 
component of the quasi-particle momentum. Note that physically 
equ.~(\ref{e_a_lag}) is a relaxation time equation for the viscous stresses. 
To demonstrate the relation to the NS equation we solve equ.~(\ref{e_a_lag}) 
for $\Delta P_a$ order by order in the small parameter ${\it Kn}=(\eta/P)
\vec{\nabla}\cdot \vec{u}$ \footnote{In kinetic theory $\eta=\tau_0 P$, 
where $\tau_0$ is the collision time, and this parameter is the Knudsen 
number ${\it Kn}$ of the flow. In fluid dynamics ${\it Kn}={\it Re}^{-1}
{\it Ma}^2$, where ${\it Re}$ is the Reynolds number, and ${\it Ma}$ is 
the Mach number.}. 
At leading order we find $\delta\Pi_{ij}=-\eta\sigma_{ij}$ and, thus, 
recover NS theory~\cite{Bluhm:2015raa}. This is true for any functional 
form of the shear viscosity $\eta(n,T)$. In the opposite limit, 
${\it Kn} \gg 1$, the components of ${\cal E}_a$  are independently 
conserved. This corresponds to the ballistic limit, because without 
collisions the components of the internal energy corresponding to motion 
in different directions are individually conserved. 

\begin{table}[t]
\begin{tabular}{|r||r|r|r|r|l|}
\hline
$T/T_F$  &  $\omega_x/(2\pi)$ & $\omega_y/(2\pi)$ & $\omega_z/(2\pi)$
         &  $\omega^{\it mag}_z/(2\pi)$            & $N$  \\ \hline\hline 
0.79     &      5283 Hz      &    5052  Hz       &  182.7 Hz
         &      21.5 Hz      &    $4\cdot 10^5$         \\ \hline
1.11     &      5283 Hz      &    5052  Hz       &  182.7 Hz
         &      21.5 Hz      &    $5\cdot 10^5$         \\ \hline
1.54     &      5283 Hz      &    5052  Hz       &  182.7 Hz
         &      21.5 Hz      &    $6\cdot 10^5$         \\ \hline

\end{tabular}
\caption{\label{tab_par}
Parameters for the experiments reported in \cite{Cao:2010wa}. The Fermi
temperature is defined in terms of the geometric mean $\bar{\omega}=
(\omega_x\omega_y\omega_z)^{1/3}$ of the trap frequencies, $T_F=(3N)^{1/3}
\bar\omega$. After the optical trap is turned off, the gas expands in a 
magnetic bowl with frequencies $\omega_i^{\it mag}$. The effect of 
$\omega_{x,y}^{\it mag}$ is negligible, and only  $\omega_{z}^{\it mag}$ is 
given in the table.}
\end{table}

{\it Comparison to solutions of the Boltzmann equation:}
Anisotropic fluid dynamics can be viewed as a low density regulator for 
the NS equation. The theory exactly reduces to NS theory in a dense fluid, 
and the relaxation time equation ensures that in the dilute limit free 
streaming is recovered. Given that the crossover between these limits is smooth
\cite{Menotti:2002,Pedri:2003,Dusling:2011dq} we expect that anisotropic 
fluid dynamics provides an accurate representation of kinetic theory at 
finite Knudsen number. Here we will verify this expectation by comparing 
numerical solutions of anisotropic fluid dynamics and the Boltzmann equation. 
The Boltzmann equation reads 
\be
\label{be}
\left( \partial_t + \vec{v}\cdot\vec{\nabla}_x 
                  - \vec{F}\cdot\vec{\nabla}_p \right) 
  f_p(\vec{x},t) = C[f_p]\, , 
\ee 
where $f_p(\vec{x},t)$ is the distribution function, $\vec{v}=\vec\nabla_p
E_p$ is the quasi-particle velocity, $E_p$ is the quasi-particle energy, 
$\vec{F}=-\vec\nabla_x E_p$ is a force, and $C[f_p]$ is the collision 
term. For simplicity, we have assumed the system to be spin-symmetric 
with $f_p^\uparrow=f_p^\downarrow=f_p$. In the high-$T$ limit $E_p=p^2/(2m)$ 
and $\vec{v}=\vec{p}/m$ \cite{Schaefer:2013oba}. In this limit the 
collision term is dominated by two-body collisions and 
\be 
 C[f_1] = -\prod_{i=2,3,4}\Big(\int d\Gamma_{i}\Big) w(1,2;3,4)
   \left( f_1f_2-f_3f_4\right)\, , 
\ee
where $f_i=f_{p_i}$, $d\Gamma_{i}=\frac{d^3p_i}{(2\pi)^3}$ 
and the transition rate is given by
\be
w(1,2;3,4) = (2\pi)^4\delta\Big(\sum_i E_i\Big)
         \delta\Big(\sum_i \vec{p}_i\Big) \,|{\cal A}|^2\, . 
\ee  
The square of the scattering amplitude in the unitary limit is given 
by $|{\cal A}|^2 = 16\pi^2/(q^2m^2)$  where $2\vec{q}=\vec{p}_2-\vec{p}_1$.
Numerical solutions of the Boltzmann equation for the unitary Fermi gas 
using the test particle method were obtained 
in~\cite{Lepers:2010,Pantel:2014jfa}. In the test particle method the 
distribution function is represented by a sum of delta functions, which 
can be thought of as classical particles that follow trajectories 
governed by Newton's laws. Collisions occur when the particles 
approach to within the scaled geometrical cross section, where the 
scale factor is determined by the number of test particles. 

 A comparison between numerical results of anisotropic fluid dynamics and 
solutions of the Boltzmann equation is shown in Fig.~\ref{fig_comp} for 
three different values of $T$~\footnote{The solutions of the Boltzmann 
equation shown in Fig.~\ref{fig_comp} correspond to results of 
\cite{Pantel:2014jfa} with all quantum corrections and in-medium effects 
removed. In the temperature-regime considered here, these effects are small.}. 
The parameters are given in Table \ref{tab_par}. The solutions of the 
Boltzmann equation were obtained for the cross section in the unitary limit.
The shear viscosity in the anisotropic fluid dynamics code is determined
by using this cross section together with the Chapman-Enskog method for 
solving the Boltzmann equation in approximate local equilibrium. The 
result of this calculation is $\eta=\frac{15}{32\sqrt{\pi}}(mT)^{3/2}$ 
\footnote{The Chapman-Enskog result is formally 
exact in the limit $n/(mT)^{3/2}\to 0$. The coefficient $\frac{15}{32
\sqrt{\pi}}$ is an approximation that arises at leading order in an 
expansion of the solution of the Boltzmann equation in Laguerre polynomials. 
The next-to-leading order correction gives a result which is larger by
a factor $193/190$ \cite{Bruun:2006,Schaefer:2014xma}. This is a 2\% 
correction, suggesting that the full result is well approximated, and 
that the correction to Fig.~\ref{fig_comp} is very small.}.
The agreement between anisotropic fluid dynamics and the Boltzmann
equation is essentially perfect. This is remarkable, because
there are no free parameters, and, as explained in the introduction, 
there are no well-behaved solutions of the NS equation 
for $\eta\sim (mT)^{3/2}$. 

 One way to think about this is to note that
it is possible to represent the Boltzmann equation as an infinite 
set of moment equations. Standard fluid dynamics corresponds to 
truncating this expansion after the first five moments, corresponding
to the conserved quantities particle number, momentum, and energy. 
What we have demonstrated is that adding only two additional moments,
corresponding to anisotropic components of the internal energy, 
dramatically improves the agreement between local moment equations
and the underlying kinetic theory for a fluid in which the density
varies significantly.

\begin{figure}[t]
\bc\includegraphics[width=7.5cm]{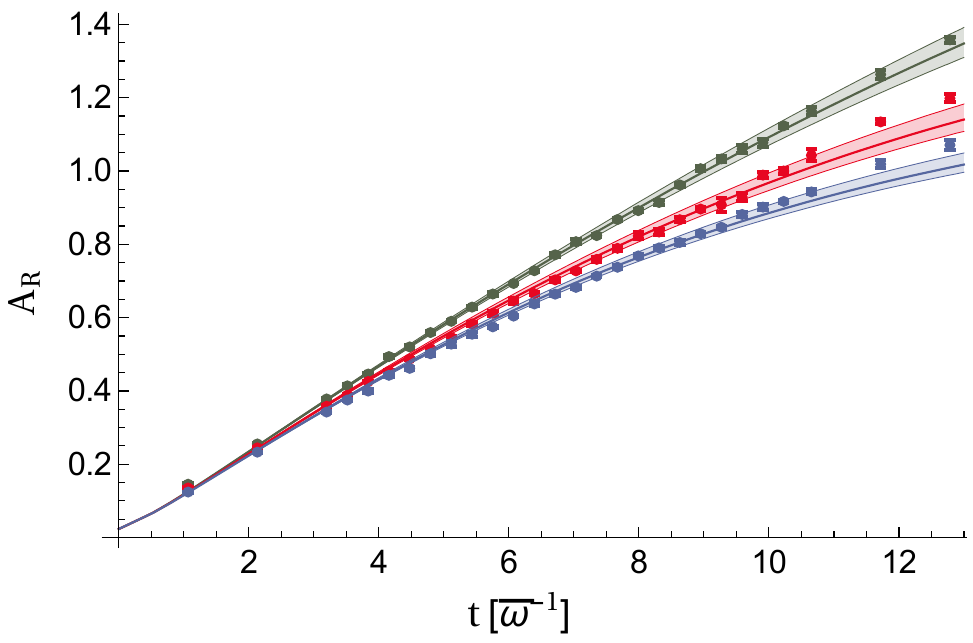}\ec
\caption{\label{fig_Cao}
 This figure shows the aspect ratio $A_R$ of an expanding unitary 
Fermi gas as a function of time $t$ in units of $\bar\omega^{-1}$ for three 
different initial temperatures $T/T_F=0.79,1.11,1.54$ (from top to 
bottom). The data are from Cao et al.~\cite{Cao:2010wa}. The solid 
lines show fits obtained using anisotropic fluid dynamics, and the 
bands correspond to a $\pm 15\%$ uncertainty in the shear viscosity.
The aspect ratio is defined using a Gaussian fit to two-dimensional
densities.}
\end{figure}

{\it Fits to high temperature expansion data:}
 We have shown that anisotropic fluid dynamics reproduces the NS 
equation in the short mean free path limit, and kinetic theory with 
two-body scattering in the dilute limit. The only input parameters are 
the equation of state and the shear viscosity. This approach is therefore 
ideally suited to determine the shear viscosity of the unitary Fermi 
gas. In this section we illustrate the method by reanalyzing the data 
of Cao et al.~\cite{Cao:2010wa}. Cao et al.~studied the expansion of the 
cloud for a range of energies $2.3N E_F\leq E\leq 4.6 NE_F$, where $E$
is the total energy of the cloud and $E_F=T_F$ is the Fermi energy. This 
is significantly above the critical energy $E_c=0.7NE_F$ for the superfluid 
transition \cite{Ku:2011}, and we can describe the initial density profile as 
a Gaussian \cite{Schafer:2010dv}. We will also assume that the shear viscosity 
follows the high temperature law $\eta=\eta_0 (mT)^{3/2}$. We will check 
this assumption below. Our goal here is to demonstrate that we can accurately 
extract the high temperature shear viscosity from data. This result provides a 
crucial and indispensable benchmark for any attempt to reliably extract the 
shear viscosity near $T_c$. 

 Fits to the data based on anisotropic fluid dynamics are shown in 
Fig.~\ref{fig_Cao}. We consider three different initial temperatures, 
spanning about a factor of two. As noted in \cite{Pantel:2014jfa}
an important ingredient in obtaining a good fit to the data is to 
follow the experimental procedure and determine the aspect ratio from 
a Gaussian fit to the two-dimensional column density $n(x,z)=\int dy\, 
n(x,y,z)$. Note that the need to perform a Gaussian fit is related to 
viscous effects. In ideal fluid dynamics the evolution preserves the 
Gaussian shape of the initial density distribution, and there is no 
difference between rms and Gaussian fit radii. At $T/T_F=0.79,1.11,1.54$ 
we find $\eta_0=0.266,0.302,0.288$. The fits to the data for these values 
of $\eta_0$, together with $\pm 15\%$ error bands, are shown in 
Fig.~\ref{fig_Cao}. There are some discrepancies at large $t$, but this 
is the regime in which systematic errors in the measurement of the
aspect ratio are expected to be significant \footnote{The goodness of 
fit $\chi^2/\nu$ for the fits shown in Fig.~\ref{fig_Cao} is between 
5 and 10. This is comparable to the quality of the model-dependent 
fits in the original work \cite{Cao:2010wa}. Note that the error bars 
in the data only include statistical fluctuations in the measurement 
of the aspect ratio.}. 

 We observe that as the temperature of the cloud changes by a factor 
of 1.95, and the shear viscosity changes by a factor 2.72, the variance of 
the extracted values of $\eta_0$ is only 6\%. This places strong constraints
on deviations from the expected scaling behavior $\eta\sim T^{3/2}$. 
Combining all the data, and using a fit to the more general functional 
form  $\eta=\eta_0(mT)^{3/2}(mT/n^{2/3})^a$, we find $a=0.05\pm 0.1$, 
consistent with $a=0$ \footnote{The best fit is $\eta=\eta_0(mT)^{3/2}
(T/T^*)^a$ with $\eta_0=0.282$, $a=0.05$ and $T^*=10.1\,T_F$. Here, $T_F
=k_F^2/(2m)$ is the local Fermi temperature. We have fixed the dimensionless 
coefficient $\eta_0$ from the fit for $a=0$.}. 
For $a=0$ we obtain $\eta=0.282(mT)^{3/2}$, which agrees to about 
5\% with the theoretical prediction $\eta=0.269(mT)^{3/2}$ 
\cite{Bruun:2005,Bruun:2006,Schaefer:2014xma}. We note that the
theoretical uncertainty inherent in the use of anisotropic fluid
dynamics, which can be estimated from Fig.~1, is much smaller than 
that. Indeed, the difference between theory and experiment is 
consistent with the statistical uncertainty of the fit, which is 
about 10\%.

{\it Conclusions and outlook:}
 In this work we have demonstrated that anisotropic fluid dynamics
can be used to make high precision, model-independent, determinations 
of the shear viscosity of trapped atomic Fermi gases. The key 
feature of the method is that it interpolates between an 
exact realization of the Navier-Stokes equation in the short 
mean free path limit and ballistic expansion in the long mean
free path limit. We have also shown that the method 
provides a very accurate representation of the Boltzmann 
equation in the limit of pure two-body scatterings. Together, these
results imply that the method incorporates the most general 
description of a dense fluid in the normal phase, Navier-Stokes 
fluid dynamics, and the correct theory of a dilute gas, kinetic
theory with two-body collisions.

 In this work we have focused on high temperature data and
verified the theoretical prediction for $\eta$ in this
regime. We have been able to extract, for the first time, the
shear viscosity coefficient without uncontrolled assumptions about 
dissipative effects in the dilute corona. This is a crucial 
benchmark for the natural next step, which is to reanalyze data 
near the superfluid transition \cite{Joseph:2014}. This will 
require initializing the density profile for a non-trivial equation 
of state, and extracting the full functional dependence of $\eta$ 
on $n/(mT)^{3/2}$. In order to describe the data below $T_c$ the 
method has to be extended to superfluid hydrodynamics. In principle
this is straightforward, because in terms of fluid dynamics a
superfluid can be viewed as a mixture of a normal, viscous, fluid
with an inviscid fluid.

 Finally, we emphasize that the method presented in this work
is quite general, and can be applied to a variety of physical problems. 
This includes problems in fluid dynamics which involve the expansion
into a vacuum, or large changes in the density, so that the 
Knudsen number of the flow varies by orders of magnitude. The 
basic idea of the method can also be applied to determine other transport 
coefficients, for example the spin diffusion 
constant in trapped atomic gases \cite{Sommer:2011}.

 Acknowledgments: This work was supported in parts by the US Department 
of Energy grant DE-FG02-03ER41260. We would like to thank Michael Urban 
for providing us with the data points shown in Fig.~\ref{fig_comp}.
We also thank James Joseph and John Thomas for many useful discussions,
and John Blondin for help with the VH1 code. 


\end{document}